\documentclass[12pt,preprint]{aastex}

\newcommand{\be}{\begin{equation}}
\newcommand{\ee}{\end{equation}}
\newcommand{\bea}{\begin{eqnarray}}
\newcommand{\eea}{\end{eqnarray}}

\begin{document}

\title{COSMIC RAY SCATTERING AND STREAMING IN COMPRESSIBLE MAGNETOHYDRODYNAMIC TURBULENCE}

\author{Huirong Yan and A. Lazarian}
\affil{Department of Astronomy, University of Wisconsin-Madison}

\date{\today{}}

\begin{abstract}
Recent advances in understanding of magnetohydrodynamic (MHD) turbulence
call for revisions in the picture of cosmic ray transport.
In this paper we use recently obtained scaling laws for MHD modes
to obtain the scattering frequency for cosmic rays. 
Using quasilinear theory we
 calculate gyroresonance with MHD modes (Alfv\'{e}nic, slow and
fast) and transit-time damping (TTD) by fast modes. We provide calculations
of cosmic ray scattering for various phases of interstellar medium
with realistic interstellar turbulence driving that is consistent
with the velocity dispersions observed in diffuse gas. We account
for the turbulence cutoff arising from both collisional and collisionless
damping. We obtain analytical expressions for diffusion coefficients that 
enter Fokker-Planck equation describing cosmic ray evolution. We obtain the scattering rate and show that fast modes
provide the dominant contribution to cosmic ray scattering for the
typical interstellar conditions in spite of the fact that fast modes
are subjected to damping. We determine how the efficiency of the scattering
depends on the characteristics of ionized media, e.g. plasma $\beta$. 
We calculate the range of energies for which
the streaming instability is suppressed by the ambient MHD turbulence. 
\end{abstract}


\section{Introduction}

Most astrophysical systems, e.g. accretion disks, stellar winds, the
interstellar medium (ISM) and intercluster medium are turbulent with
an embedded magnetic field that influences almost all of their properties.
High conductivity of the astrophysical fluids makes the magnetic fields
{}``frozen in'', and influence fluid motions. The coupled motion
of magnetic field and conducting fluid holds the key to many astrophysical
processes. 

The propagation of cosmic rays (CRs) is affected by their interaction
with magnetic field. This field is turbulent and therefore, the resonant
interaction of cosmic rays with MHD turbulence has been discussed
by many authors as the principal mechanism to scatter and isotropize
cosmic rays (Schlickeiser 2002). Although cosmic ray diffusion can
happen while cosmic rays follow wandering magnetic fields (Jokipii
1966), the acceleration of cosmic rays requires efficient scattering.
For instance, scattering of cosmic rays back into the shock is a
vital component of the first order Fermi acceleration (see Longair
1997).

While most investigations are restricted to Alfv\'{e}n modes propagating
along an external magnetic field (the so-called slab model of Alfv\'{e}nic
turbulence), obliquely propagating MHD modes have been included in
Fisk et al. (1974) and later studies (  Pryadko
\& Petrosian 1999). A more complex models were obtained by combining the 
results of the Reduced MHD with parallel slab-like
modes have been also considered (Bieber et al. 1988).
Here we attempt to use models that are motivated by
the recent studies of MHD turbulence 
(Goldreich\& Sridhar 1995, see Cho, Lazarian \& Yan 2002 for a 
review and references therein).  

A recent study (Lerche \& Schlickeiser 2001) found a strong dependence
of scattering on turbulence anisotropy. Therefore the calculations
of CR scattering must be done using a realistic MHD turbulence model.
An important attempt in this direction was carried out in Chandran
(2000). However, only incompressible motions were considered. On the 
contrary, ISM
is highly compressible. Compressible MHD turbulence has been studied
recently (see review by Cho \& Lazarian 2003a and references therein).
Schlickeiser \& Miller (1998) addressed the scattering by fast modes.
But they did not consider the damping, which is essential for fast
modes. In this paper we discuss in detail various damping processes
which can affect the fast modes. To characterize the turbulence we
use the statistics of Alfv\'{e}nic modes obtained in Cho, Lazarian
\& Vishniac (2002, henceforth CLV02) and compressible modes obtained
in Cho \& Lazarian (2002, henceforth CL02, 2003b, henceforth CL03). 

As MHD turbulence is an extremely complex process, all theoretical
constructions should be tested thoroughly. Until very recently matching
of observations with theoretical constructions used to be the only way
of testing. Indeed, it is very dangerous to do MHD turbulence testing
without fully 3D MHD simulations with a distinct inertial range.
Theoretical advances related to the anisotropy of MHD turbulence
and its scalings (see Shebalin et al. 1983, Higdon 1984, Montgomery,
Brawn \& Matthaeus 1987, Shebalin \& Montgomery 1988, Zank \& Matthaeus
1992) were mostly done in relation with the observations of fluctuations
in outer heliosphere and solar wind. Computers allowed an alternative
important way of dealing with the problem. While still presenting a limited
inertial range, they allow to control the input parameters well making it
easier to test theoretical ideas. For instance, one of us involved in
the compressible turbulence research, failed so far to observe in the
simulations the characteristic features, e.g. the production of a substantial
amount of the quasi-parallel modes that are predicted by some of earlier
MHD turbulence theories and were used to justify the slab geometry
approximation for cosmic ray scattering.  In this respect, the
observational evidence in
favor of such modes could be interpreted as the result of angle-dependent
damping that we touch upon in this paper.
                                                                                
While actual MHD turbulence is and will be in the near future the subject
of intensive research and scientific disputes, in the present paper we
adopt a simple model of turbulence that seems to be consistent
with present day numerical MHD simulations.

Yan \& Lazarian (2002, henceforth YL02) used recent advances in understanding of MHD
turbulence (CL02) to describe cosmic ray propagation in the galactic
halo. In this paper we undertake a detailed study of cosmic ray 
scattering rates in different idealized phases of the interstellar medium. 
In $\S$2,
we describe our model of turbulence, introduce 
 the statistics of Alfv\'{e}nic and compressible turbulence
in various conditions, including both high $\beta$ and low $\beta$
cases. In $\S$3, we describe the resonant interactions between the
MHD modes and CRs. The scattering
by Alfv\'{e}n modes and fast modes is presented in $\S$4. We
apply our results to different phases of ISM, including Galactic halo, warm ionized medium (WIM), hot ionized medium (HIM), and also partially
ionized medium. In $\S$5, we describe the streaming instability in
the presence of background turbulence. Discussion of our results is provided in $\S$6 while the summary is given in $\S$7.

\section{MHD cascade and its damping}

\subsection{Adopted model of MHD turbulence cascade}

A shortcoming of many earlier studies was that the crucial 
element for cosmic ray scattering, namely, MHD turbulence %%model was taken somewhat ad hoc. Our paper is an attempt to improve the situation. In what follows we
briefly describe the model of turbulent cascade adopted and refer the reader
to the original papers for a detailed explanation.

Understanding of the properties of MHD turbulence is a challenging problem
dealt with by many researchers. This work
 is not the appropriate place to cite all the
important papers that shaped the field. For instance,
a very incomplete list of such
references in a review by Cho, Lazarian \& Vishniac (2003) 
contains more than a hundred
entries.

It is well known that linear 
MHD perturbations can be decomposed into Alfv\'{e}nic, slow and fast
waves with well-defined dispersion relations 
(see Alfv\'{e}n \& F\"{a}lthammar 1963). 
Important questions arise. Can the MHD perturbations that characterize turbulence be separated into distinct modes? Can the linear modes be used for this 
purpose? Is it possible to talk about, for instance, an Alfv\'enic cascade
in a compressible medium? The answer to these non trivial 
questions is critical to studying cosmic
ray scattering.

The separation into Alfv\'en and pseudo-Alfv\'en modes, which
are the incompressible limit of Alfv\'en modes, is an essential element  
of the Goldreich-Sridhar (1995, henceforth GS95) model of turbulence. 
The arguments there were provided in favor of Alfv\'en modes
developing a cascade on their own, while the pseudo-Alfv\'en modes would
passively follow the cascade. The drain of Alfv\'enic mode energy to
pseudo-Alfv\'en modes was shown to be marginal along the cascade.
 
This model and the legitimacy of the separation into modes 
were tested successfully with incompressible MHD numerical simulations 
(Cho \& Vishniac 2000; Maron \& Goldreich 2001, CLV02). This motivated
a further study of the issue within the compressible medium.

The separation of 
MHD perturbations in compressible media into different modes 
was discussed further in Lithwick \& Goldreich 2001 and CL02. There
arguments were provided why one does not expect the modes to get
entangled into an inseparable mess even in spite of the fact
that MHD turbulence is a highly non-linear phenomenon.

The actual decomposition of MHD turbulence into Alfv\'en, slow and fast 
modes was a challenge that was addressed in CL02, CL03. 
There a particular realization of turbulence
with mean magnetic field comparable to the fluctuating magnetic field 
was studied. This setting is probably typical of the Galactic magnetic field and would allow us to use a statistical procedure of
decomposition in the Fourier space, where the basis of the Alfv\'en, slow
and fast perturbations was defined. For some particular cases, e.                                                                                                                                                                                                                                                                   g. for the case of low $\beta$ medium, the procedure was benchmarked successfully%
\footnote{In this case, the velocity of slow modes is nearly parallel to the local magnetic field, and therefore the decomposition in the real space as compared to the Fourier case decomposition is possible. The two decompositions provided identical results (CL03)} and therefore argued to be reliable
 (see CL03 for more details).

We also want to stress that the numerical studies in CL02 and CL03 were not 
blind empirical studies. They were aimed to test theoretical expectations.
The very fact that for all the tested cases the theory was
confirmed, makes us confident about the theory.  

Unlike hydrodynamic turbulence, Alfv\'{e}nic one is anisotropic, with 
eddies elongated
along the magnetic field (see Higdon 1984, Shebalin et al 1983). On the
intuitive level it can be explained as the result of the following fact:
it is easier to mix the magnetic field lines perpendicular
to the direction of the magnetic field rather than to bend them. 
However, one cannot do mixing in the perpendicular direction to very
small scales without affecting the parallel scales. This is probably the
major difference between the adopted model of Alfv\'enic perturbations and
the Reduced MHD (see Bieber et al. 1994). In the GS95 model as well as in
its generalizations for compressible medium mixing motions induce
the reductions of the scales of the parallel perturbations. 

The corresponding
scaling can be easily obtained. For instance, calculations in CLV02
prove that motions perpendicular to magnetic field lines are essentially
hydrodynamic. As the result, energy transfer rate due to those motions
is constant $\dot{E_{k}}\sim v_{k}^{2}/\tau_{k}$, where $\tau_{k}$
is the energy eddy turnover time $\sim(v_{k}k_{\perp})^{-1}$, where
$k_{\perp}$ is the perpendicular component of the wave vector $\mathbf{k}$.
The mixing motions couple to the wave-like motions parallel to magnetic
field giving a critical balance condition, i.e., $k_{\bot}v_{k}\sim k_{\parallel}V_{A}$,
where $k_{\parallel}$ is the parallel component of the wave vector
$\mathbf{k}$, $V_{A}$ is the Alfv\'en speed%
\footnote{note that the linear dispersion relation is used for Alfv\'en modes.
}. From these arguments,
the scale dependent anisotropy $k_{\parallel}\propto k_{\perp}^{2/3}$and
a Kolmogorov-like spectrum for the perpendicular motions 
$v_{k}\propto k^{-1/3}$ can be obtained (see Lazarian \& Vishniac 1999).

It was conjectured in Lithwick \& Goldreich (2001) that GS95 scaling
should be approximately true for Alfv\'{e}n and slow modes in moderately
compressible plasma. For magnetically dominated, the so-called low
$\beta$ plasma, CL02 showed that the coupling of Alfv\'{e}nic and
compressible modes is weak and that the Alfv\'{e}nic and slow modes
follow the GS95 spectrum. This is consistent with the analysis of
HI velocity statistics (Lazarian \& Pogosyan 2000, Stanimirovic \&
Lazarian 2001) as well as with the electron density statistics (see
Armstrong, Rickett \& Spangler 1995). Calculations in  CL03 demonstrated that 
fast modes are marginally
affected by Alfv\'{e}n modes and follow acoustic cascade in both
high and low $\beta$ medium. In what follows, we consider both Alfv\'{e}n
modes and compressible modes and use the description of those modes
obtained in CL02, CL03 to study CR scattering by MHD turbulence.

The distribution of energy between compressible and incompressible modes 
depends, in general, on the way turbulence is driven. 
Both the scale of driving and how the driving is performed matter.
CL02 and CL03 studied generation of compressible perturbations using random 
incompressible driving with the energy coming from the largest
scales of the system. While in the interstellar medium the energy can be
injected at different scales, the bulk of the injected energy comes from
the large scales (see Mac Low 2002) which explains the power spectra
of turbulence observed (see discussion in Lazarian \& Cho 2003). If the
injection of energy happens at velocity higher that the Alfv\'en velocity
up to the scale at which $V_l\approx V_A$ the turbulence has hydrodynamic
character. At smaller scales Alfv\'enic, fast and slow modes cascades 
develop. At the large scale the turbulence is isotropic in this case
for all modes. The opposite case of injection velocity less than 
$V_A$ results in the Alfv\'en and slow cascade being anisotropic even
at the injection scale, provided that the driving is isotropic.

How much energy is injected into fast modes if the driving takes place
at large scales?  If energy is initially injected in terms of Alfv\'enic perturbation, then the energy in fast 
$\sim\delta V_f^2$ and Alfv\'en $\sim\delta V_A^2$ modes are related by (CL02),
\be
(\delta V_f/\delta V_A)^2\sim \delta V_A\times V_A/(V_A^2+C_S^2),
\ee
where $C_S$ is the sound speed. This relation testifies that at large scales 
incompressible driving can transfer an appreciable part of energy into fast 
modes. However, at smaller scales the drain of energy from Alfv\'en to fast 
modes is marginal. Therefore the cascades evolve without much of cross talk. 
In high $\beta$ medium, however, the velocity perturbation $\delta V$ at 
injection scale may be larger than Alfv\'en speed. The injection speed of 
fast modes in this case can be approximated as 
$\delta V_f/\delta V\sim (\delta V/C_S)^2$. 
Naturally a more systematic study of different types of 
driving is required. In the absence of this, in what follows we assume that 
equal amounts of energy are transfered into fast and Alfv\'en modes when 
driving is at large scales.

\subsection{Plasma effects and damping of turbulence}

In many earlier papers Alfv\'{e}nic turbulence
was considered by many authors as the default model of interstellar
magnetic turbulence. This was partially motivated by the fact that
unlike compressible modes, the Alfv\'{e}n ones are essentially free
of damping in fully ionized medium (see Ginzburg 1961, Kulsrud \&
Pearce 1969)\footnote{This picture contradicts to an erroneous assumption
of strong coupling of compressible and incompressible MHD modes
that still percolates the literature on turbulent star formation (see
discussion in Lazarian \& Cho 2003).
However, little cross talk between the different astrophysical
communities allowed these two different pictures to coexist peacefully.}. 
In the paper below we will show that compressible fast modes are
particularly important for cosmic ray scattering. For them damping
is essential.

At small scales turbulence spectrum is altered by damping. Various
processes can damp the MHD motions (see Appendix A for details).
In partially ionized plasma, the ion-neutral collisions are the dominant
damping process. In fully ionized plasma, there are basically two
kinds of damping: collisional or collisionless damping. Their relative
importance depends on the mean free path 
in the medium (Braginskii 1965),

\be
l_{mfp}=v_{th}\tau=6\times10^{11}{\rm cm}(T/8000{\rm K})^{2}/(n/{\rm cm}^{-3}).\label{meanfp}
\ee
If the wavelength is larger than
the mean free path, viscous damping dominates. If, on the other hand,
the wavelength is smaller than mean free path, then the plasma is
in the collisionless regime and collisionless damping is dominant.

To obtain the truncation scale, the damping time $\Gamma_{d}^{-1}$
should be compared to the cascading time $\tau_{k}$. As we mentioned
earlier, the Alfv\'{e}nic turbulence cascades over one eddy turn
over time $(k_{\perp}v_{k})^{-1}\sim(k_{\parallel}V_{A})^{-1}$. The
cascade of fast modes takes a bit longer: 

\be
\tau_{k}=\omega/k^{2}v_{k}^{2}=(k/L)^{-1/2}\times V_{ph}/V^{2},\label{fdecay}
\ee
where $V$ is the turbulence velocity at the injection scale, $V_{ph}$ is is the phase speed of fast modes and equal to Alfv\'en and sound velocity for high and
low $\beta$ plasma, respectively (CL02). If the damping is faster than
the turbulence cascade, the turbulence gets
 truncated. Otherwise, for the sake
of simplicity, we ignore the damping and assume that the turbulence
cascade is unaffected. As the transfer of energy between
Alfv\'en, slow and fast modes of MHD turbulence is suppressed at the
scales less than the injection scale, we consider different components of MHD cascade independently.

We get the cutoff scale $k_{c}$ by equating the damping rate and
cascading rate $\tau_{k}\Gamma_{d}\simeq1$. Then we check whether
it is self-consistent by comparing the $k_{c}$ with the relevant
scales, e.g., injection scale, mean free path and the ion gyro-scale.

Damping is, in general, anisotropic, i.e., the linear damping (see Appendix~A) depends on the
angle $\theta$ between the wave vector ${\bf k}$ and local direction of
magnetic field ${\bf B}$. Unless randomization of $\theta$ is comparable to the cascading rate the damping scale gets angle-dependent. The angle $\theta$ varies because of both the randomization of wave vector $\bf k$ and the wandering of magnetic field lines.  Consider fast
modes, that will be shown to be 
the most important for cosmic ray scattering. Their non-linear 
cascading can be characterized by interacting 
wave vectors ${\bf k}$ that are nearly
collinear (see review by Cho, Lazarian \& Vishniac 2003). The
possible transversal deviation $\delta k$ can be estimated 
from the uncertainty condition $\delta \omega t_{cas}\sim 1$,
where $\delta \omega\sim V_{ph} \delta k (\delta k/k)$. Combining with Eq.(3), we can get 

\be
\delta k/k\simeq \frac{1}{(kL)^{1/4}}\left(\frac{V}{V_{ph}}\right)^{1/2}.
\label{dk}
\ee 
The field line wandering is mainly caused by shearing via Alfv\'en modes. The deviation of field lines during one cascading period of fast modes can be estimated as: 

\be
\frac{\delta B}{B}\simeq \left(\frac{V}{L}\right)^{1/2}\tau_k^{1/2}=\frac{1}{(kL)^{1/4}}\left(\frac{V_{ph}}{V}\right)^{1/2}.
\label{dB}
\ee
Thus the variation of the angle between ${\bf k}$ and $\bf B$ $\delta \theta\sim (1/kL)^{1/4}$ is marginal at large $k$. In the presence
of anisotropic damping this results in anisotropic distribution of
fast mode energy at large $k$.
 
With this input at hand, it is possible to determine the turbulence
damping scales for a given medium. For waves in fully ionized medium
one should compare the wavelength and mean free path before determining
which damping should be applied. However, when
 we deal with the turbulence
cascade the situation is more complicated: If the mean
free path is larger than the turbulence injection scale, we can simply
apply the collisionless damping to the whole inertial range of turbulence.
In general one should compare the viscous cut off and the mean free
path. If the cutoff scale is larger than the mean free path, it shows
that the turbulence is indeed cut off by the viscous damping. Otherwise,
we neglect the viscous damping and just apply the collisionless damping
to the turbulence below the mean free path%
\footnote{We shall show that for some angles between ${\bf B}$ and ${\bf k}$ the damping may result in turbulence transferring to collisionless regime only over a limit range of angles.%
}. By comparing different
damping, we find the dominant damping processes for the idealized
ISM phases (see Table1).
\begin{table*}
\begin{tabular}{|c|c|c|c|c|c|c|}
\hline 
ISM&
 galactic halo&
 HIM&
 WIM&
 WNM&
 CNM&
 DC\tabularnewline
\hline
T(K)&
 $2\times 10^6$&
 $1\times10^{6}$&
 8000&
 6000&
 100&
 15\tabularnewline
\hline
$C_S$(km/s)&
130&
91&
8.1&
7&
0.91&
0.35\tabularnewline
\hline
n(cm$^{-3}$)&
 $10^{-3}$&
 $4\times10^{-3}$&
 0.1&
 0.4&
 30&
 200\tabularnewline
\hline
$l_{mfp}$(cm)&
$4\times 10^{19}$&
$2\times10^{18}$&
$6\times10^{12}$&
$8\times10^{11}$&
$3\times10^{6}$&
$10^{4}$\tabularnewline
\hline
L(pc)&
 100&
 100&
 50&
 50&
 50&
 50\tabularnewline
\hline
B($\mu$G)&
5&
2&
5&
5&
5&
15\tabularnewline
\hline
$\beta$&
0.28&
3.5&
0.11&
0.33&
0.42&
0.046\tabularnewline
\hline
damping&
 collisionless&
 collisional&
 collisional&
 neutral-ion&
 neutral-ion&
 neutral-ion\tabularnewline
\hline
\end{tabular}

\caption{The parameters of idealized ISM phases and relevant damping. The
dominant damping mechanism for turbulence is given in the last line. HIM=hot ionized medium, CNM=cold neutral medium, WNM=warm neutral
medium, WIM=warm ionized medium, DC=dark cloud.}
\end{table*}

\subsection{Statistics of fluctuations}

Within random-phase approximation, the correlation tensor in Fourier
space is (see Schlickeiser \& Achatz 1993)

\begin{eqnarray}
<B_{i}(\mathbf{k})B_{j}^{*}(\mathbf{k'})>/B_{0}^{2}=\delta(\mathbf{k}-\mathbf{k'})M_{ij}(\mathbf{k}),\nonumber \\
<v_{i}(\mathbf{k})B_{j}^{*}(\mathbf{k'})>/V_{A}B_{0}=\delta(\mathbf{k}-\mathbf{k'})C_{ij}(\mathbf{k}),\nonumber \\
<v_{i}(\mathbf{k})v_{j}^{*}(\mathbf{k'})>/V_{A}^{2}=\delta(\mathbf{k}-\mathbf{k'})K_{ij}(\mathbf{k}),\end{eqnarray}
 where $B_i$, $v_i$ are respectively the
magnetic and velocity perturbation associated with the turbulence.
For the balanced cascade we consider, i.e., equal intensity of forward
and backward modes, $C_{ij}(\mathbf{k})=0$.

The isotropic tensor usually used in the literature is \begin{equation}
M_{ij}(\mathbf{k})=C_{0}\{\delta_{ij}-k_{i}k_{j}/k^{2}\} k^{-11/3},\label{isotropic}\end{equation}
 The normalization constant $C_{0}$ here can be obtained if the energy
input at the scale $L$ is defined. Assuming equipartition $u_{k}=\int dk^{3}\sum_{i=1}^{3}M_{ii}B_{0}^{2}/8\pi\sim B_{0}^{2}/8\pi$,
we get $C_{0}=L^{-2/3}/12\pi$. The normalization for the 
tensors below are obtained in the same way.

The analytical fit to the anisotropic tensor for Alfv\'{e}n modes,
obtained in CLV02 is,

\begin{equation}
\left[\begin{array}{c}
M_{ij}({\mathbf{k}})\\
K_{ij}({\mathbf{k}})\end{array}\right]=\frac{L^{-1/3}}{6\pi}I_{ij}k_{\perp}^{-10/3}\exp(-L^{1/3}|k_{\parallel}|/k_{\perp}^{2/3}),\label{anisotropic}\end{equation}
 where $I_{ij}=\{\delta_{ij}-k_{i}k_{j}/k^{2}\}$ is a 2D tensor in
$x-y$ plane which is perpendicular to the magnetic field, $L$ is
the injection scale, $V$ is the velocity at the injection scale.
Slow modes are passive and similar to Alfv\'{e}n modes. 

According to CL02, fast modes are isotropic and have one dimensional
energy spectrum $E(k)\propto k^{-3/2}$. In low $\beta$ medium, the
corresponding correlation is (YL02)\begin{equation}
\left[\begin{array}{c}
M_{ij}({\mathbf{k}})\\
K_{ij}({\mathbf{k}})\end{array}\right]={\frac{L^{-1/2}}{8\pi}}H_{ij}k^{-7/2}\left[\begin{array}{c}
\cos^{2}\theta\\
1\end{array}\right],\label{lbtensor}\end{equation}
 where $\theta$ is the angle between $\mathbf{k}$ and $\mathbf{B}$,
$H_{ij}=k_{i}k_{j}/k_{\perp}^{2}$ is also a 2D tensor in $x-y$ plane%
\footnote{Here we give only the x,y components for the perturbation since only they are used in the following calculations. The complete 3D tensors are given in Appendix B. The tensor is not formally a tensor of an isotropic vector field because the vectors of
fast waves are directed in a particular way in respect to vectors
B and k. By definition, the velocity vectors of fast modes are in
the B and k plane. This is why the tensors given here are different from eq.(\ref{isotropic})}.
The factor $\cos^{2}\theta$ represents the projection as magnetic
perturbations are perpendicular to $\mathbf{k}$. 

In high $\beta$ medium, fast modes in this regime are essentially
sound modes compressing magnetic field (GS95,
Lithwick \& Goldreich 2001, CL03). The compression of magnetic field
depends on plasma $\beta$. The corresponding x-y components of the
tensors are \begin{equation}
\left[\begin{array}{c}
M_{ij}({\mathbf{k}})\\
K_{ij}({\mathbf{k}})\end{array}\right]={\frac{L^{-1/2}}{2\pi}}\sin^{2}\theta H_{ij}k^{-7/2}\left[\begin{array}{c}
\cos^{2}\theta/\beta\\
1\end{array}\right].\label{hbtensor}\end{equation}
The velocity perturbations in high $\beta$ medium are radial, i.e., along $\mathbf{k}$, thus we have the factor $\sin^{2}\theta$ and also $V_{A}^{2}/C_{s}^{2}=2/\beta$
from the magnetic field being frozen $\omega\delta\mathbf{B}\sim\mathbf{k}\times(\mathbf{v}_{k}\times\mathbf{B})$. 
We use these statistics to calculate
cosmic ray scattering arising from MHD turbulence.

\section{Interactions between turbulence and particles}

Basically there are
two types of resonant interactions: gyroresonance acceleration
and transit acceleration (henceforth TTD). The resonant condition is $\omega-k_{\parallel}v\mu=n\Omega$ ($n=0, \pm1,2...$),
where $\omega$ is the wave frequency, $\Omega=\Omega_{0}/\gamma$
is the gyrofrequency of relativistic particle, $\mu=\cos\xi$,
where $\xi$ is the pitch angle of particles. TTD formally corresponds to $n=0$ and it requires compressible perturbations.  

We employ quasi-linear theory (QLT) to obtain our estimates. If mean magnetic field is larger than the fluctuations at the injection scale, we may say that the QLT treatment we employ defines parallel diffusion.
Obtained by applying the QLT to the collisionless Boltzmann-Vlasov
equation, the Fokker-Planck equation is generally used to describe
the evolvement of the gyrophase-averaged distribution function $f$,

\[
\frac{\partial f}{\partial t}=\frac{\partial}{\partial\mu}\left(D_{\mu\mu}\frac{\partial f}{\partial\mu}+D_{\mu p}\frac{\partial f}{\partial p}\right)+\frac{1}{p^{2}}\frac{\partial}{\partial p}\left[p^{2}\left(D_{\mu p}\frac{\partial f}{\partial\mu}+D_{pp}\frac{\partial f}{\partial p}\right)\right],\]
 where $p$ is the particle momentum. The Fokker-Planck coefficients
$D_{\mu\mu},D_{\mu p},D_{pp}$ are the fundamental physical parameter
for measuring the stochastic interactions, which are determined by
the electromagnetic fluctuations (Schlickeiser \& Achatz1993):

Adopting the approach in (Schlickeiser \& Achatz 1993) and taking into account only the dominant interaction at $n=\pm1$, we can get the Fokker-Planck
coefficients (see Appendix B),

\begin{eqnarray}
\left(\begin{array}{c}
D_{\mu\mu}\\
D_{pp}\end{array}\right)  =  {\frac{\pi\Omega^{2}(1-\mu^{2})}{2}}\int_{\bf k_{min}}^{\bf k_{max}}dk^3\delta(k_{\parallel}v_{\parallel}-\omega \pm \Omega)\nonumber\\
\left(\begin{array}{c}
\left(1+\frac{\mu V_{ph}}{v\zeta}\right)^{2}\\
m^{2}V_{A}^{2}\end{array}\right)\left\{ (J_{2}^{2}({\frac{k_{\perp}v_{\perp}}{\Omega}})+J_{0}^{2}({\frac{k_{\perp}v_{\perp}}{\Omega}}))\right.\nonumber\\
\left[\begin{array}{c}
M_{{\mathcal{RR}}}({\mathbf{k}})+M_{{\mathcal{LL}}}({\mathbf{k}})\\
K_{{\mathcal{RR}}}({\mathbf{k}})+K_{{\mathcal{LL}}}({\mathbf{k}})\end{array}\right]-2J_{2}({\frac{k_{\perp}v_{\perp}}{\Omega}})J_{0}({\frac{k_{\perp}v_{\perp}}{\Omega}})\nonumber\\
\left.\left[e^{i2\phi}\left[\begin{array}{c}
M_{{\mathcal{RL}}}({\mathbf{k}})\\
K_{{\mathcal{RL}}}({\mathbf{k}})\end{array}\right]+e^{-i2\phi}\left[\begin{array}{c}
M_{{\mathcal{LR}}}({\mathbf{k}})\\
K_{{\mathcal{LR}}}({\mathbf{k}})\end{array}\right]\right]\right\} ,\label{gyro}
\end{eqnarray}
where $\zeta=1$ for Alfv\'{e}n modes and $\zeta=k_{\parallel}/k$
for fast modes, $k_{min}=L^{-1}$, $k_{max}=\Omega_{0}/v_{th}$
corresponds to the dissipation scale, $m=\gamma m_{H}$ is the relativistic
mass of the proton, $v_{\perp}$ is the particle's velocity component
perpendicular to $\mathbf{B}_{0}$, $\phi=\arctan(k_{y}/k_{x}),$
${\mathcal{L}},{\mathcal{R}}=(x\pm iy)/\sqrt{2}$ represent left and
right hand polarization.

The delta function $\delta(k_{\parallel}v_{\parallel}-\omega+n\Omega)$
approximation to real interaction is true when magnetic perturbations
can be considered static%
\footnote{Cosmic rays have such high velocities that the slow variation of the
magnetic field with time can be neglected. %
} (Schlickeiser 1993). For cosmic rays, $k_{\parallel}v_{\parallel}\gg\omega$
so that the resonant condition is essentially $k_{\parallel}v\mu-n\Omega=0$.
From this resonance condition, we know that the most important interaction
occurs at $k_{\parallel}=k_{\parallel,res}=\Omega/v_{\parallel}$.
This is generally true except for small $\mu$ (or scattering near
$90^{o}$). 

\section{Scattering of cosmic rays}

\subsection{Scattering by Alfv\'{e}nic turbulence}

As we discussed in $\S$2, Alfv\'{e}n modes are anisotropic, eddies
are elongated along the magnetic field, i.e., $k_{\perp}>k_{\parallel}$.
The scattering of CRs by Alfv\'{e}n modes is suppressed first because
most turbulent energy goes to $k_{\perp}$ due to the anisotropy of
the Alfv\'{e}nic turbulence so that there is much less energy left
in the resonance point $k_{\parallel,res}=\Omega/v_{\parallel}\sim r_{L}^{-1}$.
Furthermore, $k_{\perp}\gg k_{\parallel}$ means $k_{\perp}\gg r_{L}^{-1}$
so that cosmic ray particles have to be interacting with lots of eddies
in one gyro period. This random walk substantially decreases the scattering
efficiency. The scattering by Alfv\'en modes was studied in YL02, (see Appendix B). In case that the pitch angle $\xi$ not close to 0, the analytical result is  \begin{equation}
\left[\begin{array}{c}
D_{\mu\mu}\\
D_{pp}\end{array}\right]=\frac{v^{2.5}\mu^{5.5}}{\Omega^{1.5}L^{2.5}(1-\mu^2)^0.5}\Gamma[6.5,k_{max}^{-\frac{2}{3}}k_{\parallel,res}L^{\frac{1}{3}}]\left[\begin{array}{c}
1\\
m^{2}V_{A}^{2}\end{array}\right],\label{ana}\end{equation}
where $\Gamma[a,z]$ is the incomplete gamma function. The presence
of this gamma function in our solution makes our results orders of
magnitude larger than those%
\footnote{We compared our result with the resonant term as the nonresonant term
is spurious as noted by Chandran (2000). %
} in Chandran (2000) for the most of energies considered. However,
the scattering frequency,

\be
\nu=2D_{\mu\mu}/(1-\mu^{2}),\label{nu}
\ee
are still much smaller
than the estimates for isotropic and slab model (see YL02). As the anisotropy of the Alfv\'{e}n modes is increasing with the
decrease of scales, the interaction with Alfv\'{e}n modes becomes
more efficient for higher energy cosmic rays. When the Larmor radius
of the particle becomes comparable to the injection scale, which is
likely to be true in the shock region as well as for very high energy cosmic
rays in diffuse ISM, Alfv\'{e}n modes get important.

It's worthwhile to mention the imbalanced cascade of Alfv\'{e}n modes
(CLV02). Our basic assumption above was that Alfv\'{e}n modes were
balanced, meaning that the energy of modes propagating one way was
equal to that in opposite direction. In reality, many turbulence sources
are localized so that the modes leaving the sources are more energetic
than those coming toward the sources. The energy transfer in the imbalanced
cascade occurs at a slower rate, and the Alfv\'{e}n modes behave
more like waves. The scattering by these imbalanced Alfv\'{e}n modes could be more efficient. However, as the actual 
degree of anisotropy of imbalanced
cascade is currently uncertain, and the process will be discussed elsewhere\footnote{Preliminary results by one of us show that the inertial range
over which the degree of anisotropy of imbalanced turbulence is small
is very limited.}. 

\subsection{Scattering by fast modes}

The contribution from slow modes is no more than that by Alfv\'{e}n modes since the slow modes have the similar anisotropies and scalings.
More promising are fast modes, which are isotropic (CL02). With fast
modes there can be both gyroresonance
and transit-time damping (TTD) (Schlickeiser \& Miller 1998).

TTD happens due to the resonant
interaction with parallel magnetic mirror force.
The advantage
of TTD is that it doesn't have a distinct resonant scale associated
with it. TTD is thus an alternative to scattering of low energy CRs which
Larmor radii are below the damping scale of the fast modes. Moreover,
we shall show later that TTD can contribute substantially to cosmic ray acceleration (also
known as the second order Fermi acceleration). This can be crucial in
some circumstances, e.g., for $\gamma$ ray burst (Lazarian et al.
2003), and acceleration of charged particles (Yan \& Lazarian
2003). Different from gyroresonance, the resonance function of TTD is broadened even for CRs with small pitch angles. The formal resonance peak $k_{\parallel}/k=V_{ph}/v_{\parallel}$ favors quasi-perpendicular modes. However, these quasi-perpendicular modes cannot form an effective mirror to confine CRs because the gradient of magnetic perturbations along the mean field direction $\nabla_{\parallel}\mathbf{B}$ is small. Thus the resonance peak is weighted out and the Breit-Wigner-type resonance function should be adopted (see eq.\ref{TTD} and Schlickeiser \& Achatz 1993).  

Here we apply our analysis to the various phases of ISM.

\begin{figure}
\includegraphics[ width=0.8\columnwidth]{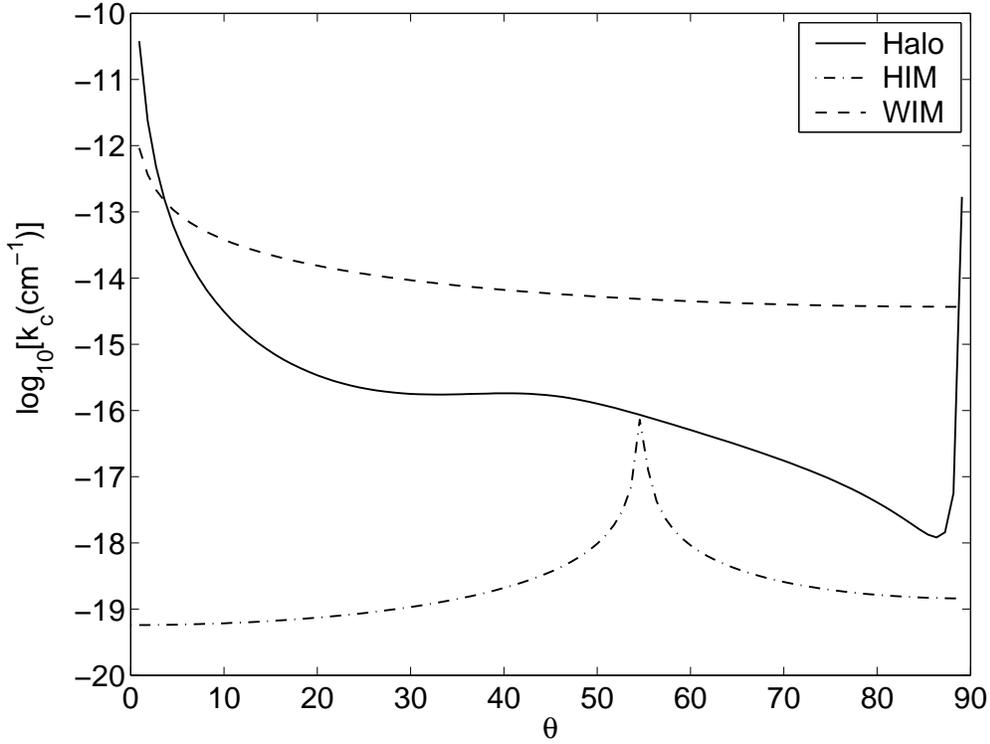}
\caption{Damping scale vs. the angle $\theta$ between ${\bf k}$ and $\bf B$ in halo, HIM and WIM.}
\end{figure}

\emph{Halo}

Recent observations (Beck 2001) suggest that the galactic halos are magnetically dominant, corresponding to low $\beta$ medium. For low $\beta$ medium, we use the tensors given by Eq.(\ref{lbtensor}). For the halo, since damping at the scale of mean free path is slower than cascading, fast modes cascade further to collisionless regime and are subjected to collisionless damping. As we see from Eq.(\ref{lbcoll}),
the damping increases with $\theta$ unless $\theta$ is close to
$\pi/2$. This
means that those fast modes in the two narrow cones with axis parallel and perpendicular to the magnetic
field are less damped. For those quasi-parallel modes the argument for the Bessel function
in Eq.(\ref{genmu}) is $k_{\perp}\tan\xi/k_{\parallel,res}<1$
unless $\xi$ is close to $90^{o}$. So we can take advantage of
the anisotropy of the damped fast modes and use the asymptotic form
of Bessel function for small argument $J_{n}(x)\simeq(x/2)^{n}/n!$
to obtain the corresponding analytical result for this case (see
Appendix~B):

\begin{eqnarray}
\left[\begin{array}{c}
D_{\mu\mu}\\
D_{pp}\end{array}\right]=\frac{\pi(\Omega v\mu)^{0.5}(1-\mu^{2})}{2L^{0.5}}\nonumber \\
\left[\begin{array}{c}
(1-(\left(\frac{k_{\perp,c}}{k_{\parallel,res}}\right)^{2}+1)^{-\frac{7}{4}})/7\\
(1-(\left(\frac{k_{\perp,c}}{k_{\parallel,res}}\right)^{2}+1)^{-\frac{3}{4}})m^{2}V_{A}^{2}/3\end{array}\right]
\label{fastlb}\end{eqnarray}
The contribution from those quasi-perpendicular modes is

\begin{eqnarray}
\left[\begin{array}{c}
\Delta D_{\mu\mu}\\
\Delta D_{pp}\end{array}\right]\simeq\frac{\pi(\Omega v\mu(1-\mu^{2}))^{0.5}}{3L^{0.5}}\nonumber \\
\left[\begin{array}{c}
\left(\frac{k_{\parallel,res}}{k'_{\perp,c}}\right)^{4.5}/4.5\\
\left(\frac{k_{\parallel,res}}{k'_{\perp,c}}\right)^{2.5}m^{2}V_{A}^{2}/2.5
\end{array}\right].
\end{eqnarray}
As we can see in Fig.1, the quasi-perpendicular modes are confined in a much narrower cone than quasi-parallel modes, which indicates that $\frac{k_{\perp,c}}{k_{\parallel,res}}\gg \frac{k_{\parallel,res}}{k'_{\perp,c}}$.  Thus $\Delta D_{\mu\mu}\ll D_{\mu\mu}$. This is understandable because quasi-perpendicular modes have the similar anisotropy as Alfv\'en modes, which has been shown negligible for CR scattering. 

From Eq.(\ref{fastlb}), we see the key factor is $(k_{\perp,c}/k_{\parallel,res})=\tan\theta_{c}$.
Using Eq.(\ref{lbcoll}) and the
relation $\tau_{k}\Gamma_{L}\simeq1$, we see the critical angle $\theta_c$ decreases with the energy of CRs $\tan\theta_{c}> 2(m_i /\pi m_e\beta k L)^{1/4}\propto r_L^{1/4}$. Therefore scattering frequency in the halo increases slowly with energy. Since the variation of $\theta$ during cascade $\delta \theta\simeq 1/(kL)^{1/4}\ll \theta_c$ at the corresponding scales, we assume the only effect of this variation is to reduce $\theta_c$ by $\delta \theta$.
Fig.2a. shows the result corresponding to a pitch angle $\xi=45^o$ and we will adopt this value for the following calculations. The contribution from TTD can be obtained by evaluating Eq.(\ref{TTD}) numerically. The contribution is mostly from quasi-perpendicular propagating modes for which the collisionless damping is minimal (see Eq.\ref{lbcoll})%
\footnote{Thus as a special case, $\delta$ function gives a good approximation in halo except for small $\mu$.}. Limited by the variation of $\theta$, the smallest scale the cascade can reach is a
scale corresponding to $k_c(\pi/2-\delta\theta)$. Combining with Eq.(\ref{dk}) and (\ref{dB}), we can solve Eq.(\ref{lbcoll}) iteratively and get $k_c(\pi/2-\delta\theta)\simeq 4.3\times 10^{-14}\rm cm^{-1}$. The contribution of TTD to scattering is smaller than that of
gyroresonance (see Fig.2a and 3a) except for small $\mu$.

\emph{WIM}

According to the results in $\S$2, the warm ionized medium (WIM)
is also in low $\beta$ but collisional regime. The mean free path $l_{mfp}=6\times10^{12}$cm
(table1), which corresponds to the resonant scale of CRs with energy
$\sim10$GeV. Thus For CRs harder than 10GeV the resonant modes with $k_{\parallel}=\Omega/v_\parallel$  are subjected to collisional damping. Since the viscous damping increases with $\theta$, we can
apply Eq.(\ref{fastlb}) to the gyroresonance in WIM as well. By equating the damping rate Eq.(\ref{viscous}) with the cascading rate Eq.(\ref{fdecay}), we can obtain $\tan\theta_{c}>V_A/(k^{3/4}l^{1/4})\sqrt{6\rho_i/\eta_0)} \propto r_L^{3/4}$. Thus according to Eq.(\ref{fastlb})
the scattering frequency curve in WIM gets steeper (see Fig.2b). The angle variation $\delta \theta\simeq 1/(kL)^{1/4} < \theta_c$ for CRs $>10$GeV, so we adopt the same treatment as above, namely, subtracting $\delta \theta$ from $\theta_c$.  For
lower energy CRs, the only available modes are those with ${\bf k}$ in a small cone as the residual of the collisional damping at large scales. Certainly, these modes are also affected by collisionless damping. However, the pitch angle of these modes is so small that the collisionless process is marginal. For scattering of CRs with low energy or large pitch angles, TTD is an alternative and can be evaluated by Eq.(\ref{TTD}) (see Fig.2b \& 3b). 

\emph{Hot Ionized Medium}

Hot ionized medium (HIM) is in high $\beta$ regime and we adopt
the tensors given by Eq.(\ref{hbtensor}). The mean free path of the
HIM is of the order of parsec and thus the fast modes in this medium
are subjected to collisional damping (see table1). We use Eq.(\ref{viscous})
to acquire the formal damping scale as a function of $\theta$. As shown in Fig.1), the peak around $\theta\sim 53^o$ indicates that damping is the weakest here. However, since the angle variation $\delta \theta\sim 1/(kL)^{1/4}\sim 20^o$ is large at the scale of mean free path $2\times 10^{18}$cm, the cascade doesn't get though mean free path. Therefore, there is no gyroresonance interaction for moderate energy CRs. The only available interaction is TTD. The scattering frequency is $\sim 1.2\times10^{-13}\rm s^{-1}$ for the energy range considered.

\emph{Partially ionized media}

Partially ionized media are similar to HIM in the sense that fast modes are severely damped by
the ion-neutral damping. The cascade is cut off before the resonant 
scales of most of the CRs we consider. Indeed, for the parameters chosen,
we find that gyroresonance only contribute to the transport of CRs
of energy $\geq0.8$TeV in DC and $>1$TeV for WNM and CNM. Therefore
only TTD contributes to the scattering of moderate energy CRs in these
media. As expected, the
scattering is less efficient in partially ionized media. The scattering frequencies for moderate energy CRs are $2.5\times 10^{-14}\rm s^{-1}$ in WNM, $\sim 4.0 \times 10^{-15}\rm s^{-1}$ in CNM and $5.7\times 10^{-15}\rm s^{-1}$ in MC.

\begin{figure}
\leavevmode
\includegraphics[ width=0.48\columnwidth]{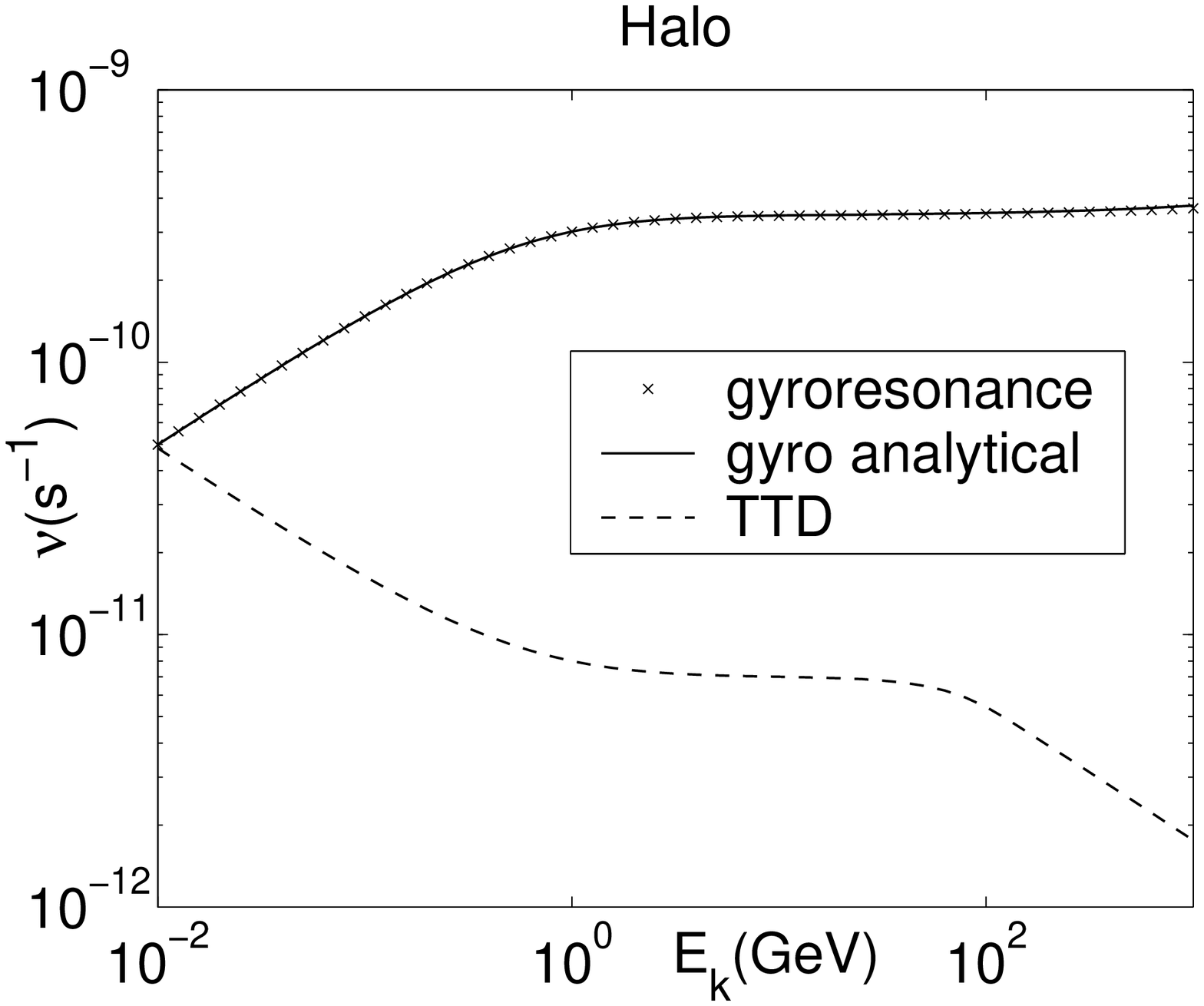} \hfil
\includegraphics[  width=0.48\columnwidth]{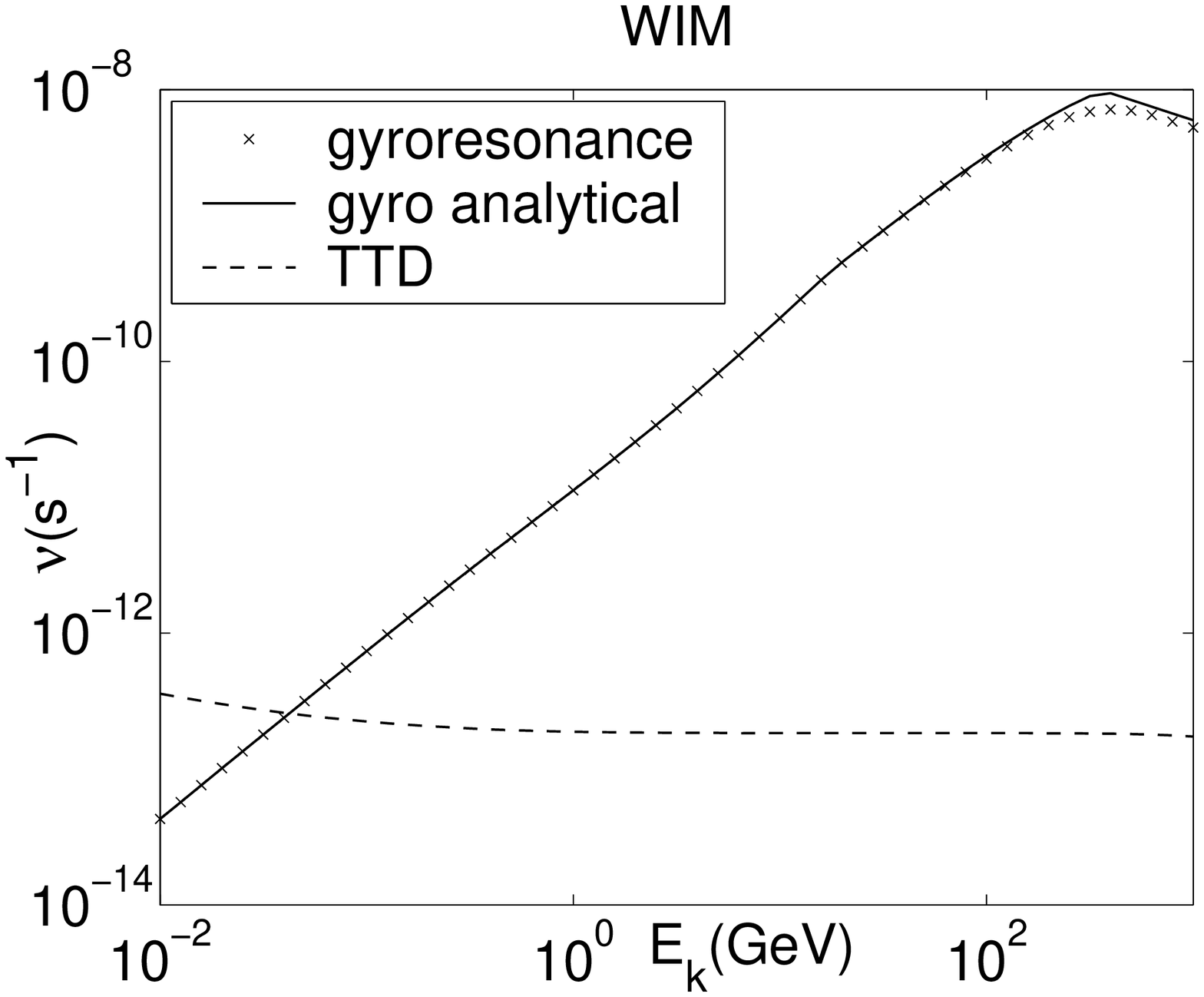} 
\caption{Scattering frequency $\nu$ given by Eq.(\ref{nu}) vs. the kinetic energy $E_{k}$ of cosmic rays (a) in halo,  (b) in WIM. The 'x' lines refer to scattering by gyroresonance and solid lines are the corresponding analytical results given by Eq.(\ref{fastlb}). 
The dashed line are the results from Eq.(\ref{TTD}) for TTD.}
\end{figure}

From the above results, we see that the fast modes dominate
CRs' scattering in spite of the damping. From Eq.(\ref{TTD}), we see the ratio of scattering and momentum diffusion
rates $p^{2}D_{\mu\mu}/D_{pp}\sim 1/\mu^{2}$, which means
that the scattering and acceleration by TTD are comparable. However,
for gyroresonance, we know that $p^{2}D_{\mu\mu}/D_{pp}\sim v^{2}/(min\{\beta,1\}V_{A}^{2})$
(see Eq.(\ref{fastlb})). Consequently for momentum
diffusion (or acceleration) TTD is dominant . 

A special case is that the cosmic rays propagate nearly perpendicular
to the magnetic field, so called the $90^{o}$ scattering problem.
It should be noted that with resonance broadening associated with the   turbulence, the requirement $v_{\parallel}=V_{ph}/\cos\theta>V_{ph}$ is relieved. The contribution from TTD thus becomes dominant for sufficiently small $\mu$. However, since quasi-linear approximation is not accurate in this regime, proper calculation should be carried out with non-linear effects taken into account (Owen 1974, Goldstein 1976).

\begin{figure}
\includegraphics[ width=0.48\columnwidth]{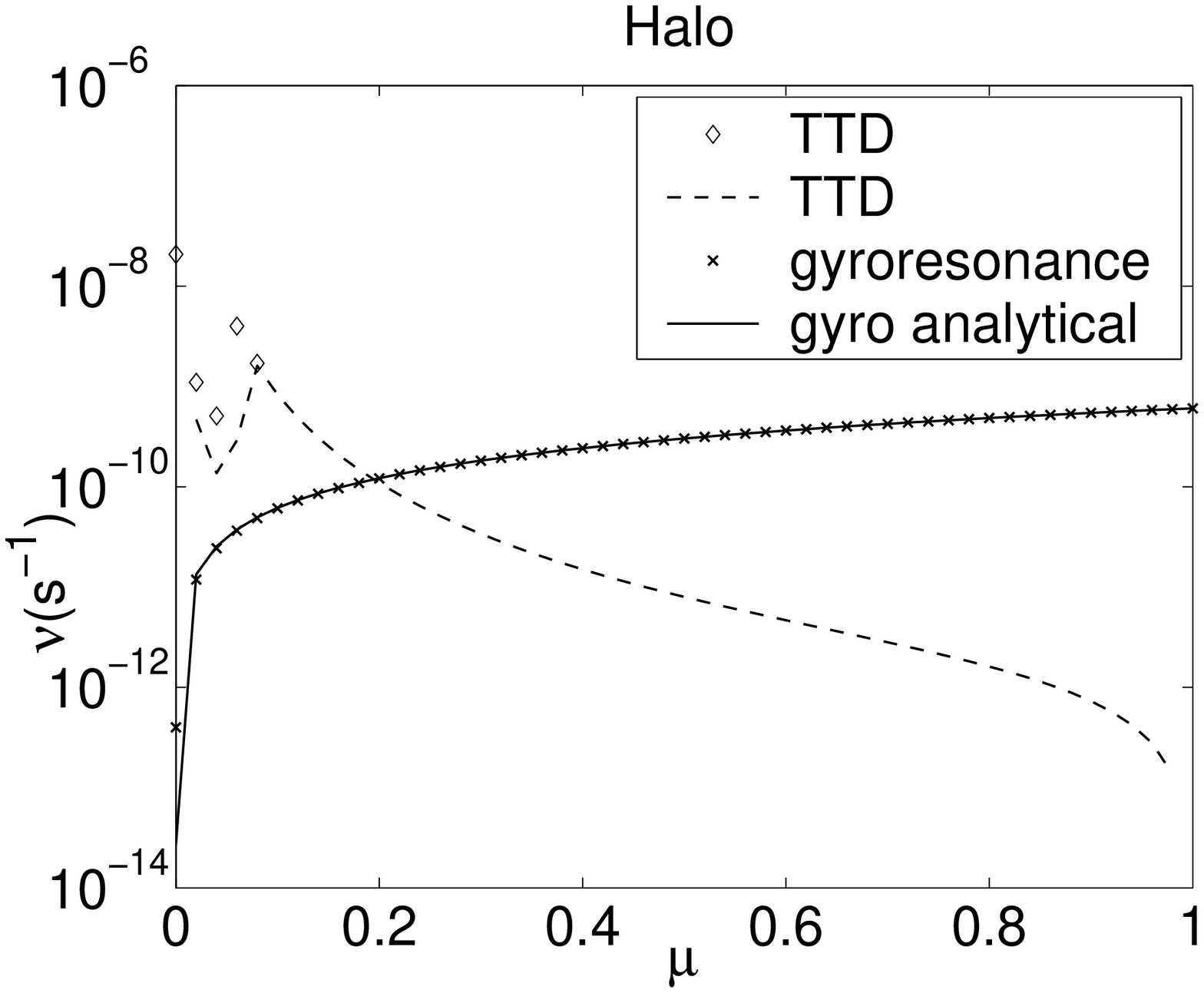} \hfil 
\includegraphics[ width=0.48\columnwidth]{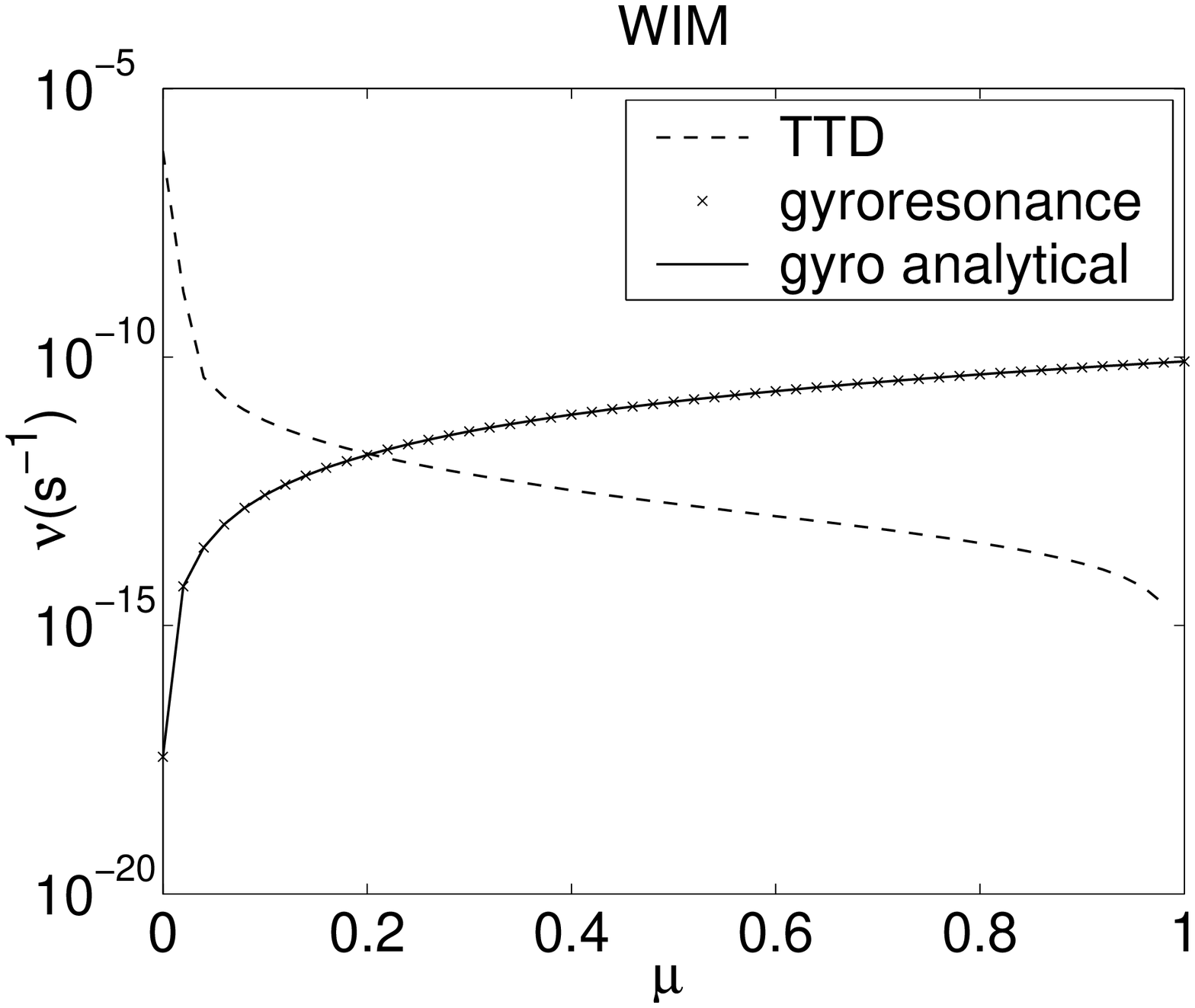} 
\caption{The scattering frequency $\nu$ vs. cosine of pitch angle $\mu$ of 1GeV CR (a) in halo, (b) in WIM.  The 'x' lines refer to scattering by gyroresonance and solid lines are the corresponding analytical results given by Eq.(\ref{fastlb}). 
The dashed lines are the results from Eq.(\ref{TTD}) for TTD. In (a), $\delta$ function and Breit-Weigner function match well except for the part marked by diamond symbols showing that $\delta$ function underestimate the contribution (see text).} 
\end{figure}

It should be pointed out that all the results above are very much
dependent on the plasma $\beta$, which is somewhat uncertain. But
the basic conclusion is definite: fast modes dominate cosmic ray
scattering except in very high $\beta$ medium where fast modes are
severely damped. This demands a substantial revision of cosmic ray
acceleration/propagation theories, and many related problems should be revisited.

\section{Cosmic ray self confinement by streaming instability}

When cosmic rays stream at a velocity much larger than Alfv\'{e}n
velocity, they can excite by gyroresonance MHD modes which in turn scatter
cosmic rays back, thus increasing the amplitude of the resonant mode. This 
runaway process is known as streaming instability. It was claimed
that the instability could provide confinement for cosmic rays with
energy less than $\sim 10^2$GeV (Cesarsky 1980). However, this was calculated
in an ideal regime, namely, there was no background MHD turbulence.
In other words, it was thought that the
self-excited modes would not be appreciably damped in fully ionized gas. 

This is not true for turbulent medium, however. YL02
pointed out that the streaming instability is partially suppressed
in the presence of background turbulence (see more in Lazarian,
Cho \& Yan 2003). More recently, detailed calculations of the 
streaming instability in the presence of background Alfv\'enic turbulence 
were presented in Farmer \& Goldreich (2003), where 
the growth rate of the modes of wave number $k$ was assumed to be (Longair
1997)

\begin{mathletters}
\begin{equation}
\Gamma(k)=\Omega_0\frac{N(\geq E)}{n_{p}}(-1+\frac{v_{stream}}{V_{A}}),
\label{instability}
\end{equation}
 where $N(\geq E)$ is the number density of cosmic rays with energy
$\geq E$ which resonate with the wave,
$n_{p}$ is the number density of charged particles in the medium.
The number density of cosmic rays near the sun is $N(\geq E)\simeq 2\times10^{-10}(E/$GeV$)^{-1.6}$cm$^{-3}$sr$^{-1}$ (Wentzel 1974).

Consider interaction with fast modes, which were shown earlier to
dominate scattering. 
Such an interaction happens at the rate 
$\tau_k\sim (k/L)^{-1/2}V_{ph}/V^2$ (see Eq.(\ref{fdecay})). 
By equating the growth rate Eq.((\ref{instability})) and the damping rate 
Eq.(\ref{fdecay}), we can find that the streaming instability
is only applicable for particles with energy less than
\begin{equation}
\gamma_{max}\simeq1.5\times 10^{-9}[n_{p}^{-1}(V_{ph}/V)(Lv\Omega_0/V^2)^{0.5}]^{1/1.1},
\end{equation}
which for HIM, provides $\sim 20$GeV if taking the injection speed to be $V\simeq 25$km/s. 

One of the most vital cases for streaming instability is that of cosmic
ray acceleration in strong MHD shocks. Such shocks produced by supernovae 
explosions scatter cosmic rays by the postshock turbulence and by preshock
magnetic perturbations created by cosmic ray streaming. The perturbations
of the magnetic field may be substantially larger than the regular magnetic
field. In this situation a different expression for the growth rate is 
applicable (Ptuskin \& Zirakashvili 2003):
\begin{equation}
\Gamma(k)\simeq\frac{ak \epsilon U^3}{c V_A(\gamma_{max}^a-(1+a)^{-1})(kr_{g0})^a(1+A_{tot}^2)^{(1-a)/2}},
\label{shock}
\end{equation}
where $\epsilon$ is the ratio of the pressure of CRs at the shock and the upstream momentum flux entering the shock front, $U$ is the shock front speed, $a-4$ is the spectrum index of CRs at the shock front, $r_{g0}=c/\Omega_0$, $A_{tot}$ is the dimensionless amplitude of the random field $A_{tot}^2=\delta B^2/B_0^2$. 

Magnetic field itself is likely to be amplified through an inverse 
cascade of magnetic energy at which perturbations created at a particular
$k$ diffuse in $k$ space to smaller $k$ thus inducing inverse cascade. 
As the result the magnetic perturbations at smaller $k$ get larger than the 
regular field. As the result, even if the instability is suppressed
for the growth rate given by eq. (\ref{instability}) it gets efficient
due to the increase of perturbations of magnetic field stemming from the
inverse cascade.

Whether or not the streaming instability is efficient in scattering
accelerated cosmic rays back depends on whether the growth rate of
the streaming instability is larger or smaller than the damping rate.
The precise picture of the process depends on yet not completely
clear details of the inverse cascade of magnetic field. 
If, however, we assume that the small scale driving provides at 
the scales of interest isotropic turbulence the nonlinear damping
happens on the scale of one eddy turnover time. If assuming the shock front speed $U$ is low enough, then the advection of  the turbulence by shock flow is negligible and the perturbation $A_{tot}\ll 1$. In this case, we can get the maximum energy of particles accelerated in the shock by equating the growth rate of the instability Eq.(\ref{shock}) to the damping rate due to turbulence cascade Eq.(\ref{fdecay}): 
\begin{equation}
\gamma_{max}\simeq (\frac{a\epsilon (LeB_0)^{0.5}U^3}{m^{0.5}V^{2}c^2})^{1/(0.5+a)}.
\end{equation}
From this we can estimate $\gamma_{max}\simeq 2\times 10^7 (t/kyr)^{-9/4}$ in HIM.

As shown in the Eq.(\ref{instability}) and (\ref{shock}), the
growth rate depends on the CRs' density. In those regions where high
energy particles are produced, e.g., shock fronts in ISM, $\gamma$ ray burst, SN, the
streaming instability is more important.

\end{mathletters}

\section{Discussion}

\subsection{Applicability of our results}

We have attempted to apply a new scaling relations obtained for
MHD turbulence to the problem of cosmic ray scattering taking into
account the turbulence cutoff arising from both collisional and collisionless
damping. We considered both gyroresonance
and transit-time damping (TTD). 

In earlier papers it was frequently 
assumed that Alfv\'enic turbulence follows Kolmogorov 
spectrum while
the problem of CR scattering was considered. This is incorrect because
MHD turbulence is different from hydrodynamic turbulence if magnetic
field is dynamically important. We appeal to the earlier research (see review by Cho \& Lazarian 2003a) which showed that MHD turbulence can be decomposed into Alfv\'{e}n, slow
and fast modes. They have different statistics (see CL02). The Alfv\'{e}n
and slow modes follow GS95 scalings and show scale-dependent anisotropy.

Fast modes, however, are promising as a means of scattering due to their 
isotropy. Even in spite of the damping fast modes dominate scattering if 
turbulent
energy is injected at large scales. We provided calculations for various
phases of ISM, including galactic halo, WIM, HIM and partially ionized media. As the fast
modes are subjected to different damping processes, the CR 
scattering varies. In galactic halo, the fast
modes are subjected to collisionless damping. In WIM and HIM ion viscosity dominates the damping. All these damping 
increase with $\theta$ and marginal for parallel modes. As the result, the 
fast modes at small scales become anisotropic with 
$k_{\parallel}\gg k_{\perp}$. In this sense, fast modes have slab geometry but with less energies. These modes can efficiently scatter CRs through gyroresonance. TTD is more important for large pitch angle scattering and for the regime where fast modes are severely damped. We find that use of $\delta$ function entails errors and resonance  broadening ought to be taken into account.  The formal resonance peak $\cos\theta=V_{ph}/v_{\parallel}$ is smeared out because  perturbations along the mean field $\nabla_{\parallel}\mathbf{B}$ is small for the quasi-perpendicular modes. In all the cases, the scattering efficiency decreases with plasma $\beta$
because the damping increases with $\beta$. 

As every other model, the model adopted has its own limitations. 
For instance, within our model we did not consider wavelike component
(see Montgomery \& Turner 1981, Kinney \& McWilliams 1998). Such
parallel propagating fluctuations are observed in the solar wind
(Matthaeus et al. 1990) and included in some of the existing models
of cosmic ray transport (see Bieber et al. 1994). One could argue that such fluctuations
could potentially be described by the weak turbulence model (see
Galtier et al. 2000) or the weak turbulence model with imbalance
(see a special case in Lithwick \& Goldreich 2003). 

However, weak balanced turbulence is known to have a limited inertial
range: turbulence get strong very soon 
(see discussions of this point in Cho, Lazarian \& Vishniac
2003). A work by one of us also shows that the same is true for
the weak imbalanced turbulence\footnote{In addition the propagation
of the imbalanced turbulence is limited in a compressible medium
(Cho \& Lazarian, in preparation)}. Therefore, while such type
of turbulence could indeed be important near the sources of turbulence, its implications over vast spaces
of interstellar gas are limited. 

The same is true about our model of energy injection. We assume
that the turbulent energy is being injected at large scales.
While observations of power spectra are consistent with such a picture,
it is clear that some energy is being injected at small
scales. If the intensity of fluctuations produced by this local injection is higher than those created by the global
turbulent cascade, such regions can be important places for
scattering of cosmic rays. If Alfv\'enic turbulence which provides rather weak scattering
when injected from large scales (Chandran 2000, YL02)
were the only mechanism for scattering of cosmic rays, 
such localized regions of weak turbulence or
intensive local injection of turbulent energy would be the
dominant places of scattering. As we identified
fast modes as an efficient source of scattering, in the zeroth approximation
in our treatment we disregarded those localized sources of scattering. 

We disregarded shocks as sources of scattering. We believe
that shocks are more important means of cosmic ray acceleration than 
scattering. To simplify our model further we also 
disregarded magnetic mirrors created by molecular clouds (Chandran 2001),
which statistics is uncertain and seem to be important most if
other scattering means fail. 

There are other simplifications that were adopted in the paper.
For instance, our model disregarded complications  
associated with the large scale Alfv\'en perturbations. For the
range of cosmic rays that we are dealing the Alfv\'en fluctuations are
well within the inertial range and therefore various non-linearities
are not important. 
% I'd like to delete this sentence. 
Due to the same reasoning, 
we do not deal with the effects of
"bendover" associated with the energy injection scale (see Zank et al
1998). 

Similarly, we do not
deal with the gyroresonance scattering of the particles moving to 
magnetic field
at angles close to 90 degrees. This allows us not to consider
complex kinetic effects that are important for the small
scale turbulence (compare Leamon et al. 1999). 
In addition, we
did not consider effects of turbulence in partially
ionized gas below the scale at which kinetic energy fluctuations
are damped by the viscosity. This new regime of MHD turbulence
arises in the partially ionized gas (Cho, Lazarian \& Vishniac 2002)
and is characterized by the high degree of magnetic field intermittency.
While our simple estimates do not show that this regime is important
for cosmic ray scattering, effects of magnetic intermittency on cosmic ray
transport do require more work.

How good is the decomposition when the mean magnetic field is not
strong is the issue that must still be tested. The decomposition
procedure used in CL02 and CL03 requires existence of the mean
magnetic field. However,
one may argue (see CL03) that when mean magnetic field
is weak, locally the nonzero magnetic field of larger eddies acts as
the mean magnetic field for the smaller ones. Therefore physically the results obtained for the particular model could be argued to be valid even in 
the absence of the mean magnetic field. Testing this numerically
is possible by taking volumes in the real space, defining there the local
direction of the local mean field that changes from one volume to another
(see Cho, Lazarian \& Vishniac 2003 for a detailed discussion) 
and then using a decomposition in the Fourier space as this is done in
CL02. For a present numerical studies this is extremely challenging, 
however. Nevertheless, the correspondence of properties of MHD turbulence 
with and without mean magnetic field reported in Cho \&
Lazarian (2003b) 
 supports the theoretical expectations about the modes. 
One way or another, there are theoretical expectations that have been
tested for the case most important for scattering of Galactic cosmic rays.
These were being used for the present study.

Although scattering via fast modes is isotropic at large scales and becomes 
slab-type at small scales at which damping gets important,  our model is 
radically different from earlier discussed 2D+slab and 3D+slab models 
(Bieber et al. 1994, Schlickeiser \& Miller 1998).We used theoretically 
motivated models of turbulence that have been tested numerically. Therefore 
the transitions from one regime to another as well as the values of the 
intensity of perturbations have good justification within our approach.

We adopted a particular set of parameters of ISM while doing the calculation.
The basic conclusion, namely, fast modes dominate the cosmic ray scattering,
will remain the same even if the parameters are altered. However, since
the $\beta$ value in ISM remains uncertain from observation,
the damping scale $k_{c}$ is somewhat uncertain. And this determines
the energy limit down to which gyroresonance is applicable.

\subsection{Comparison with recent attempts in the same
direction}

An earlier attempt to study the effects of the GS95 scaling on the propagation
of cosmic rays is made in Chandran (2000) and YL02 . These studies showed that Alfv\'en modes
are not be efficient for cosmic ray scattering. 
Moreover, YL02 identified fast modes
as the driver for cosmic ray scattering. As fast modes are subjected to
damping it was not clear whether fast mode induced scattering is always efficient.
This problem was dealt with in the present paper, which
provides detailed calculations of the expected scattering rate by fast modes
for various phases of the ISM and takes into account both gyroresonance and
TTD scattering.

Another issue that this paper deals with is the suppression of the cosmic
ray streaming instability by turbulence. This effect was first noticed
in YL02 (see Lazarian,
Cho \& Yan (2003), Yan \& Lazarian (2003) for more detail). The
calculations of the effect were presented by Farmer \& Goldreich (2004) for
the weakly perturbed magnetic field. As the magnetic field may get strongly
perturbed due to the inverse cascade of energy as the streaming instability
develops first at smaller scales, in this paper we provide calculations
of the suppression of the streaming instability for the strongly
perturbed magnetic field.

\section{Summary}

In the paper above we characterized interaction of cosmic rays with 
balanced interstellar turbulence driven at a large scale. Our results 
can be summarized as follows:

1. Our calculations show that 
fast modes provide the dominant contribution to scattering of cosmic
rays in different phases of interstellar medium
provided that the turbulent energy is injected at large scales.
This happens in spite of the fact that the fast modes are more
subjected to damping compared to Alfv\'en modes. 

2. As damping of fast modes depends on the angle between
the magnetic field and the wave direction of propagation,
we find that field wondering determined by Alfv\'en modes affects
the damping of fast modes. At small scales the 
anisotropy of fast mode damping
makes gyroresonant scattering within slab approximation 
applicable. At larger 
scales where damping is negligible the isotropic gyroresonant 
scattering approximation is applicable.

3. Transient time damping (TTD) provides an important means of how cosmic rays can be accelerated by fast modes. Use of $\delta$ function resonance entails errors, and therefore, resonance broadening is essential for TTD. Our study shows that it is vital for low energy and large pitch angle scattering. And it dominates scattering of cosmic rays in HIM and the partially 
ionized interstellar gas where fast modes are severely damped.

4. Streaming instability is subjected to non-linear damping
 due to the interaction of 
the emerging magnetic perturbations with the surrounding turbulence.
The energy at which the streaming instability is suppressed
depends on whether on the inverse cascade of magnetic energy
as the instability gets more easily excited for low energy particles.

\begin{acknowledgments}
We acknowledge valuable discussions with Jungyeon Cho, Steve Cowley, Don Cox, Randy Jokipii, Vladimir Mirnov, and Reinhard Schlickeiser. We thank the referee for helpful comments. This work is supported by the NSF grant AST0307869 ATM 0312282 and the NSF Center
for Magnetic Self Organization in Laboratory and Astrophysical Plasmas.

\end{acknowledgments}

\appendix

\section{A. Damping of MHD turbulence}
Below we summarize the damping processes that we consider in the paper.

\emph{Ion-neutral damping}

In partially ionized medium, viscosity of neutrals provides damping
(see LY02). If the mean free path for a neutral atom, $l_{n}$, in
a partially ionized gas with density $n_{tot}=n_{n}+n_{i}$ is much
less than the size of the eddies under consideration, i.e. $l_{n}k\ll1$,
the damping time

\begin{equation}
\Gamma_{in}\sim\nu_{n}k^2\sim\left(\frac{n_{tot}}{n_{n}}\right)(l_{n}c_{n})k^{2},\label{tdamp}\end{equation}
 where $\nu_{n}$ is effective viscosity produced by neutrals%
\footnote{The viscosity due to ion-ion collisions is typically small as ion
motions are constrained by the magnetic field. %
}. The mean free path of a neutral atom $l_{n}$ is influenced both
by collisions with neutrals and with ions. The rate at which neutrals
collide with ions is proportional to the density of ions, while the
rate at which neutrals collide with other neutrals is proportional
to the density of neutrals. The drag coefficient for neutral-neutral
collisions is $\sim1.7\times10^{-10}T$(K)$^{0.3}$ cm$^{3}$ s$^{-1}$
(Spitzer 1978), while for neutral-ion collisions it is $\sim{\langle v_{r}\sigma_{in}\rangle}\approx1.9\times10^{-9}$
cm$^{3}$ s$^{-1}$ (Draine, Roberge \& Dalgarno 1983). Thus collisions
with other neutrals dominate for $n_{i}/n_{n}$ less than $\sim0.09T^{0.3}$.
Turbulent motions cascade down till the cascading time is of the order
of $t_{damp}$. The maximal damping corresponds to $k^{-1}\sim l_{n}$.
If the neutrals constitute less than approximately $5\%$, the cascade
goes below $l_{n}$ and is damped at smaller scale (see below).

\emph{Collisionless damping}

The nature of collisionless damping is closely related to the radiation
of charged particles in magnetic field. Since the charged particles
can emit plasma modes through acceleration (cyclotron radiation) and
Cherenkov effect, they also absorb the radiation under the same condition
(Ginzburg 1961). The thermal particles can be accelerated either by
the parallel electric field which can also be called Landau damping
or the magnetic mirror (TTD) associated with the comoving modes (or
under the Cherenkov condition $k_{\parallel}v_{\parallel}=\omega$).
The gyroresonance with thermal ions also causes the damping of those
modes with frequency close to the ion-cyclotron frequency (Leamon et al. 1998),  though
it is irrelevant to the low frequency modes since we deal with GeV CRs here. The
collisionless damping depends on the plasma $\beta\equiv P_{gas}/P_{mag}$
and the propagation direction of the modes. In general, the damping
increases with the plasma $\beta$. And the damping is much more severe
for fast modes than for Alfv\'{e}n modes. For instance, the Alfv\'{e}n
modes are weakly damped even in a high $\beta$ medium, where fast
modes are strongly damped. The damping rate $\gamma_{d}=\tau_{d}^{-1}$
of the fast modes of frequency $\omega$ for $\beta\ll1$ and $\theta\sim1$
(Ginzburg 1961) is

\begin{eqnarray}
\Gamma_{L} & = & \frac{\sqrt{\pi\beta}}{4}\omega\frac{\sin^{2}\theta}{\cos\theta}\times[\sqrt{\frac{m_{e}}{m_{H}}}\exp(-\frac{m_{e}}{m_{H}\beta\cos^{2}\theta})+ 5\exp(-\frac{1}{\beta\cos^{2}\theta})],
\label{lbcoll}
\end{eqnarray}
 where $m_{e}$ is the electron mass. The exact expression for the
damping of fast modes at small $\theta$ was obtained in Stepanov%
\footnote{We corrected a typo in the corresponding expression.%
} (1958)

\[
\Gamma_{L}=\frac{\sqrt{\pi\beta}}{4}\omega\theta^{2}\times\left(1+\frac{\theta^{2}}{\sqrt{\theta^{4}+4\Omega_{i}^{2}/\omega^{2}}}\right)\sqrt{\frac{m_{e}}{m_{H}}}\exp(-\frac{m_{e}}{m_{H}\beta\cos^{2}\theta}).\]
 When $\beta\gg1$, we obtain the damping rate as a function of $\theta$ from Foote \& Kulsrud 1979,

\begin{eqnarray}
\Gamma_{L}=\frac{\sin^2\theta}{\cos^3\theta}\times \cases{2\omega^{2}/\Omega_{i}  &  for $k<\Omega_{i}/\beta V_{A}$\cr
2\Omega_{i}/\beta  &  for $k>\Omega_{i}/\beta V_{A}$\cr}
\label{hbcoll}
\end{eqnarray}
 where $\Omega_{i}$ is the ion gyrofrequency.

\emph{Ion viscosity}

In a strong magnetic field ($\Omega_{i}\tau_{i}\gg1$) the transport
of transverse momentum is prohibited by the magnetic field. Thus transverse
viscosity $\eta_{\perp}$ is much smaller than longitudinal viscosity
$\eta_{0}=0.96n(T/eV)\tau_{i}$, $\eta_{\perp}\sim\eta_{0}/(\Omega_{i}\tau_{i})^{2}$.
The heat generated by this damping is $Q_{vis}=\eta_{0}(\partial v_{x}/\partial x+\partial v_{y}/\partial y-2\partial v_{z}/\partial z)^{2}/3$,
where $v_{x}$, $v_{y}$, $v_{z}$ are the velocity components of
the wave perturbation (Braginskii 1965). From the expression, we see
that the viscous damping is not important unless there is compression.
Therefore Alfv\'{e}n modes is marginally affected by the ion viscosity.

While the damping due to compression along the magnetic fields can
be easily understood, it is counterintuitive that the compression
perpendicular to the magnetic also results in damping through longitudinal
viscosity. Following Braginskii (1965), the damping of perpendicular
motion may be illustrated in the following way. To understand the
physics of such a damping, consider fast modes in low $\beta$ medium.
In this case, the motions are primarily perpendicular to the magnetic
field so that $\partial v_{x}/\partial x=\dot{n}/n\sim\dot{B}/B$.
The transverse energy of the ions increases because of the conservation
of adiabatic invariant $v_{\perp}^{2}/B$. If the rate of compression
is faster than that of collisions, the ion distribution in the momentum
space is bound to be distorted from the Maxwellian isotropic sphere
to an oblate spheroid with the long axis perpendicular to the magnetic
field. As a result, the transverse pressure gets greater than the
longitudinal pressure by a factor $\tau_{i}\sim\dot{n}/n$, resulting
in a stress $\sim p\tau_{i}\dot{n}/n\sim\eta_{0}\partial v_{x}/\partial x$.
The restoration of the equilibrium increases the entropy and causes
the dissipation of energy with a damping rate $\Gamma_{ion}=k_{\perp}^{2}\eta_{0}/6\rho_{i}$
(Braginskii 1965).

In high $\beta$ medium, the velocity perturbations are radial as
pointed out in $\S2$. Thus $Q_{vis}=\eta_{0}(k_{x}v_{x}+k_{y}v_{y}-2k_{z}v_{z})^{2}/3=\eta_{0}k^{2}v^{2}(\sin^{2}\theta-2\cos^{2}\theta)^{2}/3$.
Dividing this by the total energy associated with the fast modes $E_{k}=\rho_{i}v^{2}$,
we can obtain the damping rate $\Gamma_{ion}=k^{2}\eta_{0}(1-3\cos^{2}\theta)^{2}/6\rho_{i}$.

All in all,

\begin{eqnarray}
\Gamma_{ion}=\cases{	k_{\perp}^{2}\eta_{0}/6\rho_{i}   & for $\beta\ll 1$\cr
					k^{2}\eta_{0}(1-3\cos^{2}\theta)^{2}/6\rho_{i} &  for $\beta\gg 1$\cr}
\label{viscous}
\end{eqnarray}

\emph{Resistive damping}

In the paper we ignore the resistive damping because of the following reason. The resistivity is much
smaller than the longitudinal ion viscosity for fast modes. For parallel
propagating fast modes which are not subjected to the viscous damping
from the longitudinal viscosity, the resistive damping scale can be
obtained by equating the damping rate $c^{2}k^{2}/8\pi\sigma_{\perp}$
with the cascading rate of the fast modes, where $\sigma_{\perp}\simeq10^{13}(T/eV)^{3/2}$
(see Kulsrud \& Pearce 1969) is the conductivity perpendicular to
the magnetic field: $k_{res}^{-1}=1.8\times10^{7}(L/pc)^{1/3}(T/10^{4}K)^{-1}(V_{A}/kms^{-1})^{-2/3}$cm.
This scale turns out to be much smaller than the mean free path, where
collisionless damping takes over. For Alfv\'{e}n modes, the resistive
damping rate is $\Gamma_{res}=c^{2}k_{\perp}^{2}/8\pi\sigma_{\parallel}$
(the term associated with $k_{\parallel}$is negligible because of
anisotropy), where $\sigma_{\parallel}=1.96\sigma_{\perp}$is the
conductivity parallel to the magnetic field. By equating it with the
cascading rate of Alfv\'{e}n modes, we can get the resistive damping
scale $k_{res}^{-1}\simeq L(c^{2}/8\pi\sigma_{\parallel}V_{A}L)^{3/4}$.
Comparing with the proton Larmor radius, we find $k_{res}^{-1}/r_{L}=3.6\times10^{-3}(L/pc)^{1/4}(T/10^{4}K)^{-3/2}n^{1/2}(V_{A}/C_{s})^{1/4}$.
It's clear that in many astrophysical plasmas the resistive scale
is less than proton Larmor radius. At this scale the energy can be
transfered either to protons or to electron whistler modes. 
According to Cho \& Lazarian (2004) whistler modes are even more anisotropic
that the Alfv\'en ones. Therefore their heating of protons is unlikely.
In terms of cosmic ray scattering they are of marginal importance.

\section{B. Fokker-Planck coefficients}

In quasi-linear theory (QLT), the effect of MHD modes is studied by
calculating the first order corrections to the particle orbit in the
uniform magnetic field, and the ensemble-averaging over the statistical
properties of the MHD modes (Jokipii 1966, Schlickeiser \& Miller
1998). Obtained by applying the QLT to the collisionless Boltzmann-Vlasov
equation, the Fokker-Planck equation is generally used to describe
the evolvement of the gyrophase-averaged particle distribution. The
Fokker-Planck coefficients $D_{\mu\mu},D_{\mu p},D_{pp}$ are the
fundamental physical parameter for measuring the stochastic interactions,
which are determined by the electromagnetic fluctuations:

$\begin{array}{cc}
<B_{\alpha}(\mathbf{k})B_{\beta}^{*}(\mathbf{k'})>=\delta(\mathbf{k}-\mathbf{k'})P_{\alpha\beta}(\mathbf{k}) & <B_{\alpha}(\mathbf{k})E_{\beta}^{*}(\mathbf{k'})>=\delta(\mathbf{k}-\mathbf{k'})T_{\alpha\beta}(\mathbf{k})\\
<E_{\alpha}(\mathbf{k})B_{\beta}^{*}(\mathbf{k'})>=\delta(\mathbf{k}-\mathbf{k'})Q_{\alpha\beta}(\mathbf{k}) & <E_{\alpha}(\mathbf{k})E_{\beta}^{*}(\mathbf{k'})>=\delta(\mathbf{k}-\mathbf{k'})R_{\alpha\beta}(\mathbf{k}),\end{array}$

Adopting the approach in Schlickeiser \& Achatz (1993), we can get the Fokker-Planck coefficients,

\begin{eqnarray}
\left[\begin{array}{c}
D_{\mu\mu}\\
D_{\mu p}\\
D_{pp}\end{array}\right]&=&{\frac{\Omega^{2}(1-\mu^{2})}{2B_{0}^{2}}}{\mathcal{R}}e\sum_{n=-\infty}^{n=\infty}\int_{\bf k_{min}}^{\bf k_{max}}dk^{3}  \left[i\begin{array}{c}
\left(1+\frac{\mu V_{A}}{v\zeta}\right)^{2}\\
\left(1+\frac{\mu V_{A}}{v\zeta}\right)mc\\
m^{2}c^{2}\end{array}\right]\int_{0}^{\infty}dte^{-i(k_{\parallel}v_{\parallel}-\omega+n\Omega)t}\left\{ J_{n+1}^{2}({\frac{k_{\perp}v_{\perp}}{\Omega}})\left[\begin{array}{c}
P_{{\mathcal{RR}}}({\mathbf{k}})\\
T_{{\mathcal{RR}}}({\mathbf{k}})\\
R_{{\mathcal{RR}}}({\mathbf{k}})\end{array}\right]\right.\nonumber \\
&+&J_{n-1}^{2}({\frac{k_{\perp}v_{\perp}}{\Omega}})\left[\begin{array}{c}
P_{{\mathcal{LL}}}({\mathbf{k}})\\
-T_{{\mathcal{LL}}}({\mathbf{k}})\\
R_{{\mathcal{LL}}}({\mathbf{k}})\end{array}\right]+J_{n+1}({\frac{k_{\perp}v_{\perp}}{\Omega}})J_{n-1}({\frac{k_{\perp}v_{\perp}}{\Omega}})\left.\left[e^{i2\phi}\left[\begin{array}{c}
-P_{{\mathcal{RL}}}({\mathbf{k}})\\
T_{{\mathcal{RL}}}({\mathbf{k}})\\
R_{{\mathcal{RL}}}({\mathbf{k}})\end{array}\right]+e^{-i2\phi}\left[\begin{array}{c}
-P_{{\mathcal{LR}}}({\mathbf{k}})\\
-T_{{\mathcal{LR}}}({\mathbf{k}})\\
R_{{\mathcal{LR}}}({\mathbf{k}})\end{array}\right]\right]\right\} 
\label{genmu}\end{eqnarray}
 where $|{\bf k_{min}}|=k_{min}=L^{-1}$, $|{\bf k_{max}}|=k_{max}$ corresponds to the dissipation
scale, where $\zeta=1$ for Alfv\'{e}n modes and $\zeta=k_{\parallel}/k$
for fast modes, $dk^3=2\pi k_\perp dk_\perp dk_\parallel$, $m=\gamma m_{H}$ is the relativistic mass of the
proton, $v_{\perp}$ is the particle's velocity component perpendicular
to $\mathbf{B}_{0}$, $\phi=\arctan(k_{y}/k_{x}),$ ${\mathcal{L}},{\mathcal{R}}=(x\pm iy)/\sqrt{2}$
represent left and right hand polarization. For low frequency MHD
modes, we have from Ohm's Law $\mathbf{E}(\mathbf{k})=-(1/c)\mathbf{v}(\mathbf{k})\times\mathbf{B}_{0}$.
So we can express the electromagnetic fluctuations $T_{ij},R_{ij}$
in terms of correlation tensors $K_{ij}$. For $\omega$ we use dispersion relations $\omega=k_\parallel V_A$ for Alfv\'en modes and $\omega=kV_{ph}$ for fast modes. Those dispersion relations were used to decompose and study the evolutions of MHD modes in CL02.

For gyroresonance, the
dominant interaction is the resonance at $n=\pm 1$. By integrating over
$\tau$, we obtain

\begin{eqnarray}
\left(\begin{array}{c}
D_{\mu\mu}\\
D_{\mu p}\\
D_{pp}\end{array}\right) & = & {\frac{\pi\Omega^{2}(1-\mu^{2})}{2}}\int_{\bf k_{min}}^{\bf k_{max}}dk^3\delta(k_{\parallel}v_{\parallel}-\omega+\pm \Omega)\left(\begin{array}{c}
\left(1+\frac{\mu V_{ph}}{v\zeta}\right)^{2}\\
\left(1+\frac{\mu V_{ph}}{v\zeta}\right)mV_{A}\\
\iota^2m^{2}V_{A}^{2}\end{array}\right)\label{gyroapp}\\
 &  & \left\{ (J_{2}^{2}({\frac{k_{\perp}v_{\perp}}{\Omega}})+J_{0}^{2}({\frac{k_{\perp}v_{\perp}}{\Omega}}))\left[\begin{array}{c}
M_{{\mathcal{RR}}}({\mathbf{k}})+M_{{\mathcal{LL}}}({\mathbf{k}})\\
-C_{{\mathcal{RR}}}({\mathbf{k}})-C_{{\mathcal{LL}}}({\mathbf{k}})\\
K_{{\mathcal{RR}}}({\mathbf{k}})+K_{{\mathcal{LL}}}({\mathbf{k}})\end{array}\right]\right.\nonumber \\
 & - & 2J_{2}({\frac{k_{\perp}v_{\perp}}{\Omega}})J_{0}({\frac{k_{\perp}v_{\perp}}{\Omega}})\left.\left[e^{i2\phi}\left[\begin{array}{c}
M_{{\mathcal{RL}}}({\mathbf{k}})\\
-C_{{\mathcal{RL}}}({\mathbf{k}})\\
K_{{\mathcal{RL}}}({\mathbf{k}})\end{array}\right]+e^{-i2\phi}\left[\begin{array}{c}
M_{{\mathcal{LR}}}({\mathbf{k}})\\
-C_{{\mathcal{LR}}}({\mathbf{k}})\\
K_{{\mathcal{LR}}}({\mathbf{k}})\end{array}\right]\right]\right\},\nonumber \end{eqnarray}
where $\iota=1$ for low $\beta$ medium and $\iota=\sin^2\theta$ for high beta medium.

For TTD,

\begin{eqnarray}
\left(\begin{array}{c}
D_{\mu\mu}\\
D_{\mu p}\\
D_{pp}\end{array}\right) & = & {\frac{\pi\Omega^{2}(1-\mu^{2})}{2}}\int_{\bf k_{min}}^{\bf k_{max}}dk^3\frac{\tau_k}{\tau_k^2+(k_{\parallel}v_{\parallel}-\omega)^2}\left(\begin{array}{c}
\left(1+\frac{\mu V_{ph}}{v\zeta}\right)^{2}\\
\left(1+\frac{\mu V_{ph}}{v\zeta}\right)mV_{A}\\
\iota^2m^{2}V_{A}^{2}\end{array}\right)\label{TTD}\\
 &  & 2J_{1}^{2}({\frac{k_{\perp}v_{\perp}}{\Omega}})\left[\begin{array}{c}
M_{{\mathcal{RR}}}({\mathbf{k}})+M_{{\mathcal{LL}}}({\mathbf{k}})+e^{i2\phi}M_{{\mathcal{RL}}}({\mathbf{k}})+e^{-i2\phi}M_{{\mathcal{LR}}}({\mathbf{k}})\\
-C_{{\mathcal{RR}}}({\mathbf{k}})-C_{{\mathcal{LL}}}({\mathbf{k}})-e^{i2\phi}C_{{\mathcal{RL}}}({\mathbf{k}})-e^{-i2\phi}C_{{\mathcal{LR}}}({\mathbf{k}})\\
K_{{\mathcal{RR}}}({\mathbf{k}})+K_{{\mathcal{LL}}}({\mathbf{k}})+e^{i2\phi}K_{{\mathcal{RL}}}({\mathbf{k}})+e^{-i2\phi}K_{{\mathcal{LR}}}({\mathbf{k}})\end{array}\right]\nonumber 
\label{TTD}
\end{eqnarray}

Define the integrand 

\begin{eqnarray}
G(\mathbf{k})&= & (J_{2}^{2}({\frac{k_{\perp}v_{\perp}}{\Omega}})+J_{0}^{2}({\frac{k_{\perp}v_{\perp}}{\Omega}}))(K_{{\mathcal{RR}}}({\mathbf{k}})+K_{{\mathcal{LL}}}({\mathbf{k}}))-2J_{2}({\frac{k_{\perp}v_{\perp}}{\Omega}})J_{0}({\frac{k_{\perp}v_{\perp}}{\Omega}})(e^{i2\phi}K_{{\mathcal{RL}}}({\mathbf{k}})+e^{-i2\phi}K_{{\mathcal{LR}}}({\mathbf{k}}))\nonumber\\
T(\mathbf{k})&= & 2J_{1}^{2}({\frac{k_{\perp}v_{\perp}}{\Omega}})((K_{{\mathcal{RR}}}({\mathbf{k}})+K_{{\mathcal{LL}}}({\mathbf{k}})+(e^{i2\phi}K_{{\mathcal{RL}}}({\mathbf{k}})+e^{-i2\phi}K_{{\mathcal{LR}}}({\mathbf{k}}))
\end{eqnarray}.
We shall show below how they can be simplified in various cases to
enable an analytical evaluation of the integral. The spherical components
of the correlation tensors are obtained in the following.

For Alfv\'en modes, their tensors are proportional to

\[
I_{ij}=\left(\begin{array}{cc}
\sin^{2}\phi & -\cos\phi\sin\phi\\
-\cos\phi\sin\phi & \cos^{2}\phi\end{array}\right).\]

For Alfv\'en modes, $k_{\perp}v_{\perp}/\Omega\gg1$ because of the
anisotropy. So we can use the asymptotics of the Bessel function $J_{n}(x)=\sqrt{2/\pi x}\cos(x-n\pi/2-\pi/4)$
when $x\rightarrow\infty$. Thus from Eq.(\ref{anisotropic}),
we simplify the integrand in Eq.(\ref{gyro}), $G({\mathbf{k}})\propto k_{\perp}^{-10/3}\exp(-L^{1/3}|k_{\parallel}|/k_{\perp}^{2/3}),$
the integral of which will be $\propto(|k_{\parallel}|L^{1/3})^{-6.5}\Gamma[6.5,k_{max}^{-2/3}|k_{\parallel}|L^{1/3}]$.
Then from Eq.(\ref{gyro}) we can get the analytical result for the
scattering by Alfv\'en modes as given in Eq.(\ref{ana}).

For fast modes, their magnetic correlations have such forms

\[
M_{ij}\propto\left(\begin{array}{ccc}
\cos^2\theta\cos^{2}\phi & \cos^2\theta\cos\phi\sin\phi & \cos\theta\sin\theta\cos\phi\\
\cos^2\theta\cos\phi\sin\phi & \cos^2\theta\sin^{2}\phi & \cos\theta\sin\theta\sin\phi\\
\cos\theta\sin\theta\cos\phi & \cos\theta\sin\theta\sin\phi & \sin^2\theta \end{array}\right)\label{FRRLL}.
\]

Their velocity correlations are
\begin{eqnarray}
K_{ij}\propto\left(\begin{array}{ccc}
\cos^{2}\phi & \cos\phi\sin\phi & 0\\
\cos\phi\sin\phi & \sin^{2}\phi & 0\\
0 & 0 & 0 \end{array}\right)\label{FRL}. \end{eqnarray}
for $\beta\ll 1$ and $K_{ij}\propto k_i k_j/k^2$ for $\beta\gg 1$.

In general, it's more difficult to solve the integral in Eq.(\ref{genmu})
for the fast modes because they are isotropic. However, if taking
into account damping, Bessel function can be evaluated using the zeroth
order approximation. Thus from Eq.(\ref{gyro},\ref{FRRLL},\ref{FRL}),
we see for gyroresonance the integrand can be simplified as $G(\mathbf{k})\propto(J_{2}-J_{0})^{2}k^{-7/2}\propto k^{-7/2}$.
For TTD, $T(\mathbf{k})\propto J_{1}^{2}(k_{\perp}v_{\perp}/\Omega)k^{-7/2}\propto k^{-3/2}v_{\perp}^{2}\sin^{2}\theta\propto k^{-3/2}v_{\perp}^{2}(1-V_{ph}^{2}/v_{\parallel}^{2})$.
Then we can solve the integral in Eq.(\ref{TTD}) analytically and get the
scattering rate for fast modes.

\end{document}